\documentclass[amstex,12pt]{article}
\usepackage{amssymb}

\begin{document}

\title{The Egalitarian Sharing Rule in Provision of Public Projects}
\author{Anna Bogomolnaia\thanks{Rice University, Houston, USA.} \and Michel Le
Breton\thanks{Universit\'{e} de Toulouse I, GREMAQ and IDEI,
Toulouse, France.} \and Alexei Savvateev\thanks{Central Economics and
Mathematics Institute, Moscow; Institute for Theoretical and
Experimental Physics, Moscow; New Economic School,
Moscow; CORE, Catholic University of Louvain-la-Neuve, Belgium.
Financial support through grants R98-0631 from the Economic
Education and Research Consortium, \# NSh-1939.2003.6 School
Support, Russian Foundation for Basic Research No. 04-02-17227, and
the Russian Science Support Foundation is gratefully
acknowledged.} \and Shlomo Weber\thanks{SMU, Dallas, USA, and CORE,
Catholic University of Louvain, Belgium, and CEPR.}}
\date{}
\maketitle

\begin{abstract}
In this note we consider a society that partitions itself into
disjoint jurisdictions, each choosing a location of its public
project and a taxation scheme to finance it. The set of public
project is {\em multi-dimensional}, and their costs could vary
from jurisdiction to jurisdiction. We impose two principles,
\emph{egalitarianism}, that requires the equalization of the total
cost for all agents in the same jurisdiction, and
\emph{efficiency}, that implies the minimization of the aggregate
total cost within jurisdiction. We show that these two principles
always yield a core-stable partition but a Nash stable partition
may fail to exist.
\end{abstract}

\vspace{.5in}
\begin{description}
\item[Keywords:] Jurisdictions, stable partitions, public projects,
egalitarianism.
\end{description}

\vspace{.5in}

\begin{description}
\item[JEL Classification Numbers:] C71, C72, D63, H41.
\end{description}

\newpage \baselineskip=18pt

\section{Introduction}
\hspace{.21in} In this note we consider a model of jurisdiction
formation, where the entire population has to be partitioned into
several jurisdictions, each deciding on the public project from the
multi-dimensional characteristics set. Whereas the existing
literature deals with uni-dimensional universe of public projects,
in this note we consider a set of projects to be imbedded in the set
of an arbitrary dimension. Agents are assumed to have Euclidean
preferences over possible project locations, and, thus, can be
identified by their peaks (best preferred locations) in the
multi-dimensional space.

When a jurisdiction is formed and a public project is selected,  the
jurisdiction must choose the way to distribute the project cost
among its members. The discussion on the cost sharing methods in
this context has progressed in two directions. One group of papers
(Le Breton and Weber (2003), (2004), Haimanko et al. (2004a), Le
Breton et al. (2004)) assumes the transferable utility framework with
the unrestricted set of redistribution schemes within a
jurisdiction. However, in many situations the degree of freedom in
selecting a taxation scheme and degree of compensation to
disadvantaged agents could be limited by customs, law, feasibility
or variety of other constraints. Thus, a large number of papers
(Alesina and Spolaore (1997), Casella (2001), J\'{e}hiel and
Scotchmer (2001), Haimanko et al. (2004b)) consider an \emph{equal
share} taxation scheme where all members of jurisdiction make an
equal monetary contribution towards the projects costs. This cost
sharing mechanism has some appealing properties, such as simplicity
and anonymity. However, it rules out any degree of equalization
between agents who enjoy vastly different benefits from the public
project.

In this note we consider the case where every jurisdiction applies
\emph{egalitarian rule} of full equalization, when the total burden
of taxation cost and disutility from having a project different from
agent's ideal choice is equally divided among the members of the
jurisdiction. The egalitarian rule in this context implements the
Rawlsian principle by equalizing the total costs of all agents and,
therefore, minimizing the total cost of the most disadvantaged
agent. Under the egalitarian rule agents are not hold accountable
for their preferences and the burden of total disutility is shifted
to the entire jurisdiction.

The equal burden of all agents within every jurisdiction turns our
model into \emph{hedonic} game (Banerjee et al. (2001), Bogomolnaia
and Jackson (2002)), where once a coalition is formed, one can
uniquely determine the (equal) payoff of all of its members. Thus,
agents form well-defined preferences over possible coalitions they
could join. This feature of the hedonic model allows us to consider
not only traditional cooperative stability notions (like core), but
also non-cooperative stability, when only one agent can migrate to
another jurisdiction or create a new one. We show that whereas the
core stable partitions always exist in our multi-dimensional
framework, Nash stable partition may fail to exist even in the case
where the set of projects is represented by a uni-dimensional space.

\section{The Model}
\hspace{.21in} We consider a finite society $N=\{1,\ldots,n\}$ of
agents. The set of public projects is given by the compact subset
$I$ of $k$-dimensional Euclidean space $\Re_k$. Each agent has
Euclidean preferences over $I$, which allows us to identify an agent
$i$ with his/her ideal point $p^{i} \in I$. Slightly abusing the
terminology, we will refer to $p^{i}$ as the location of the agent $i$.

The society faces a task of partitioning itself into disjoint
jurisdictions, where each jurisdiction selects a location of the
public project and a taxation scheme to finance it. We assume
voluntary participation: the benefits from public projects exceed
their costs. When the location of the project $p$ and a taxation
scheme within jurisdiction $S$ are chosen, every member $i$ of $S$
incurs two types of cost: transportation cost from his/her location at
$p^i$ to $p$ and the monetary contribution towards provision of the
public project.

The locational and redistributional choices within each jurisdiction
are guided by two principles, \emph{egalitarianism} and
\emph{efficiency}. The egalitarian principle demands the equality of
the total cost (transportation and project contribution) for all
agents in the same jurisdiction. This requirement does, in fact,
generate the Rawlsian allocation (see Le Breton et al. (2004)), that
minimizes the total cost of the most disadvantaged agent within the
jurisdiction. The efficiency requires the minimization of the
aggregate total cost incurred by the members of the jurisdiction.

We assume that the cost of a public project in jurisdiction $S$ is
given by the positive value $g(S)$, which may depend on the size and
composition of a jurisdiction. The taxation scheme $x: S \rightarrow
\Re$ assigns the contribution $x(i)$ for each agent $i\in S.$ The
only requirement is that the allocation $x$ must satisfy the budget
constraint: $\sum_{i\in S} x(i)=g(S)$. In addition to the monetary
contribution $x(i)$, each member of $S$ incurs disutility, or
\emph{transportation cost}, $d\left(||p^i-p||\right)$, where $p$ is the
jurisdictional choice of the public project. We assume that function $d$
is continuous and increasing on $\Re$ with $d(0)=0$.

Given the location $p$ of the public project, the aggregate cost of members
of jurisdiction $S$ will be the sum of the cost of the public project $g(S)$
and the aggregate transportation cost $D(S,p)$, where
\[
D(S,p)= \sum_{i\in S} d \left
(||p-p^{i}||\right).
\]
The efficiency requires that the project location is chosen in such a way as
to minimize $D(S,p)$. Let
\[
D(S)= \min_{p \in I} D(S,p),
\]
and the location $p(S)$ (not necessarily uniquely defined) is
determined by $D(S,p(S)) =D(S)$.

Under the \emph{egalitarian taxation scheme}, all members of $S$
equally share the aggregated total cost (tax plus transportation
cost) $g(S)+D(S)$. Thus, every member $i$ of $S$ will incur the
same cost $c(S)$ defined by
\[
c(S)=\frac{\left( g+D(S)\right)}{|S|},
\]
where $|S|$ is the cardinality of jurisdiction $S$. Given that
$i$'s transportation cost is $d(||p^{i}-p(S)||)$, her tax share
(that can actually be a subsidy) should be
\[
\frac{\left( g(S) +D(S)\right)}{|S|}
-d\left(||p^{i}-p(S)||\right).
\]

Efficiency and egalitarianism completely determine the choices for
each potential jurisdiction. Thus, these requirements lead to
hedonic cooperative coalition formation game, where once a
jurisdiction is formed, the payoff or cost for each its member is
uniquely determined. We will examine the existence of stable
partitions in our hedonic game, where all jurisdictions adopt the
egalitarian rule. We consider two notions of stability, where the
first is the standard notion of the core of the coalition structure
(Aumann and Dr$\grave e$ze (1974)):
\begin{description}
\item[Definition 1:]  Let $\pi =\{S_{1},\ldots ,S_{K}\}$ be a jurisdictional
structure. We say that a jurisdiction $S\subset N$ {\em blocks} $\pi$
if $c(S^{i})>c(S)$ for all $i\in S$, where $S^{i}$ is the
jurisdiction in $\pi$ that contains $i$.

A jurisdictional structure $\pi$ is called \emph{core-stable} if no
jurisdiction blocks $\pi$.
\end{description}

The second definition is that of Nash stability:
\begin{description}
\item[Definition 2:]  A jurisdictional structure $\pi= \{S_{k}\}_{k=1,\ldots,K}$
is \emph{Nash stable} if
\[
c(S^{i})\leq g(\{i\})\;\mbox{ and }\;c(S^{i})\leq c(S_{k}\sqcup
\{i\})
\]
for for every $i\in S$ and every $S_{k}\in \pi $.
\end{description}

One can view a Nash stable jurisdictional structure as a \emph{free
mobility equilibrium}: no agent has an incentive to move from his/her
current jurisdiction to either ``empty'' jurisdiction or to another
existing jurisdiction. Note also that the set of Nash stable
jurisdictional structures is the set of pure Nash equilibria of the
non-cooperative game, where each agent announces his/her ``address''
and all agents with the same address form a jurisdiction.

While the definition of Nash stability allows only for deviations by
a single agent, an agent can move to a jurisdiction without the
consent of its members. Thus, some or even all members of that
jurisdiction could be worse off. This observation shows that there
is no logical connection between the notions of core- and Nash
stability. And, indeed, the next section indicates sharply different
stability implications with regard to these two notions.

\section{The Results}
\hspace{0.21in} In the egalitarian game, all members of the same
coalition bear the same total cost (or total disutility). Thus, each
coalition $S$ is assigned a number $c(S)$ that represents the total
cost of each of its members. Thus, all agents derive their
preferences over coalitions from the common ordering by comparing
jurisdictions on the basis of their assigned contributions. In
hedonic games, such common ordering guarantees the existence of a
stable partition. Our first proposition indeed states that, without
any further qualification, stable jurisdictional structures always
exist in the egalitarian game.
\begin{description}
\item[Proposition 1:]  The Egalitarian game always admits core-stable
jurisdictional structures.
\end{description}
{\bf Proof}\footnote{Since the egalitarian game satisfies the {\em
common ranking} property of Farrell and Scotchmer (1988), our result
could be derived from theirs. However, we opted to offer a direct
proof that also implies that if all values $c(S)$ are different, the
core-stable partition is unique.}: Let us construct the partition
$\pi$ as follows. Take a coalition $T^{1}$ that minimizes the
contribution of its members across all coalitions:
\[
c(T^{1})=\min_{S\subset N}c(S).
\]
(obviously, the choice of $T^{1}$ as well as of all subsequent elements of
$\pi$ is not necessarily unique.)

If $T^1=N$, we are done. Otherwise, choose $T^2 \subset N \setminus
T^1$ to minimize the cost over all coalitions that have an empty
intersection with $T^1$:
\[
c(T^2) = \min_{S \subset N \setminus T^1} c(S).
\]

If $T^{1}\sqcup T^{2}=N$, we are done. Otherwise, choose
$T^{3}\subset N\setminus (T^{1}\sqcup T^{2}))$ to minimize the cost
over all coalitions that have an empty intersection with $T^1 \sqcup T^{2}$:
\[
c(T^{3})=\min_{S\subset N\setminus (T^{1}\sqcup T^{2})}c(S).
\]
By continuing this process, after at most $n$ iterations, we obtain
a partition $\pi=\{T^{1},\ldots ,T^{K}\}$ of $N$. We show that $\pi$
is stable. Indeed, consider an arbitrary coalition $T \subset N$, and let
$T^{k}$ be the first coalition in $\pi$ that has a nonempty intersection
with $T$, i.e., $k=\min \{j|T^{j}\bigcap T\neq \emptyset \}$.
Then the choice of $T^k$ implies that $c(T^k)\leq c(T)$.
Thus, no agent that belongs to both $T^k$ and $T$, is better off at
$T$ as compared to $T^k$, which means that $T$ could not block $\pi$.

It is worthwhile pointing out that if all values $c(S)$ are
different, our construction yields a unique outcome $\pi$. It is
easy to demonstrate that $\pi$ is the unique core stable partition.
Indeed, $T^1$ has to be a part of every stable partition, otherwise this
very coalition will block the latter one; given that, $T^2$ must belong
to it as well. By extending the argument, we obtain the uniqueness
of the core stable partition in this case.$\Box $

While we can guarantee core stability in the egalitarian game, such
general statement does not hold for Nash stability:
\begin{description}
\item[Proposition 2:]  Under the Egalitarian rule a Nash stable
jurisdictional structure may fail to exist even if the public
project space is uni-dimensional.
\end{description}
\textbf{Proof:} Let space of public projects $I$ be given by the
uni-dimensional interval $[0,4.1]$ and consider a society with six
agents, located at points $p^{1}=0$, $p^{2}=p^{3}=1.9$,
$p^{4}=p^{5}=p^{6}=4.1$. Let transportation costs be linear, i.e.,
$d(|p,p'|)=|p-p'|$ for all $p,p' \in I$, and project costs $g(S)$
be equal to $1$ for all jurisdictions $S$. We will show that under
the Egalitarian rule, this society does not admit a Nash stable
partition. Suppose, in negation, that $\pi$ is a Nash stable
partition.  Denote $T=\{4,5,6\}$. It is easy to verify
that if there is $S \in \pi$ that contains one or two agents from
$T$ in addition to some of agents $1,2,3$, then $c(S)> 1$. Thus,
if agents $4,5,6$ are separated in at least two groups, they
cannot be joined by other agents, implying that the members of $T$
should belong  to the same jurisdiction in a Nash stable
partition.

Since the members' cost exceeds $1$ in every jurisdiction with
more than four members, and agent $1$ contributes strictly more
than one in all multi-agent coalitions, except $\{1,2,3\}$, it
follows that only candidates for Nash stable jurisdiction
structures are:\\
$\{\{1,2,3\},T\}$, $\{\{1\},\{2\},\{3\} \sqcup T \}$,
$\{\{1\},\{3\},\{2\} \sqcup T\}$, $\{\{1\},\{2,3\},T\}$.

Consider the following cycle:
\begin{displaymath}
\{\{1,2,3\}, T \} \rightarrow \{\{1,2\},\{3\} \sqcup T \}
\rightarrow \{\{1\},\{2\},\{3\} \sqcup T\}
\rightarrow\\
\{\{1\},\{2,3\},T\} \rightarrow \{\{1,2,3\},T\},
\end{displaymath}
in which every partition is being obtained from the previous one
by a shift of one of the dissatisfied agents to a jurisdiction
that yields his/her a lower cost. Thus, no partition in the cycle is
Nash stable, and due to a symmetry between agents $2$ and $3$, it
rules out all candidates to constitute a Nash stable partition.
$\Box$

To conclude, let us notice that the only core-stable partition in
this example is $\{\{1\},\{2,3\},T \}$.

\end{document}